# Hybrid Heuristic Algorithms for Adiabatic Quantum Machine Learning Models

Bahram Alidaee[a*], Haibo Wang[b], Lutfu S.Sua[c], and Wade W. Liu[d]

[a*] *Dept. of Marketing, School of Business Admin., Univ. of Mississippi, Oxford, MS, USA; balidaee@bus.olemiss.edu*

[b] *Division of International Business and Technology Studies, Texas A&M International Univ., Laredo, Texas, USA* [a] *Dept. of Marketing, School of Business*

[c] *Management & Marketing Dept., Southern Univ. and A&M College, Baton Rouge, LA, USA*

[d] *Dept. of Computer Science, School of Engineering, Univ. of Mississippi, Oxford, MS, USA*

# Hybrid Heuristic Algorithms for Adiabatic Quantum Machine Learning Models

Numerous established machine learning models and various neural network architectures can be restructured as Quadratic Unconstrained Binary Optimization (QUBO) problems. A significant challenge in Adiabatic Quantum Machine Learning (AQML) is the computational demand of the training phase. To mitigate this, approximation techniques inspired by quantum annealing, like Simulated Annealing and Multiple Start Tabu Search (MSTS), have been employed to expedite QUBO-based AQML training. This paper introduces a novel hybrid algorithm that incorporates an "r-flip" strategy. This strategy is aimed at solving large-scale QUBO problems more effectively, offering better solution quality and lower computational costs compared to existing MSTS methods. The r-flip approach has practical applications in diverse fields, including cross-docking, supply chain management, machine scheduling, and fraud detection. The paper details extensive computational experiments comparing this r-flip enhanced hybrid heuristic against a standard MSTS approach. These tests utilize both standard benchmark problems and three particularly large QUBO instances. The results indicate that the r-flip enhanced method consistently produces high-quality solutions efficiently, operating within practical time constraints.

## I. Introduction

Recent advancements in AQML techniques and their applications, particularly those leveraging the QUBO model, have garnered considerable interest from both researchers and industry professionals (Biamonte et al., 2017; Date et al., 2021; Guan et al., 2021; Hatakeyama-Sato et al., 2022; Orús et al., 2019; von Lilienfeld, 2018). Recent advancements in AQML techniques and their applications, particularly those leveraging the QUBO model, have garnered considerable interest from both researchers and industry professionals (Biamonte et al., 2017; Date et al., 2021; Guan et al., 2021; Hatakeyama-Sato et al., 2022; Orús et al., 2019; von Lilienfeld, 2018,. Ma et al., 2023). Many traditional machine learning algorithms, including linear regression (Date & Potok, 2021), support vector machines (Biamonte et al., 2017; Date et al., 2021), balanced k-means clustering (Date et al., 2021), feature subset selection (Chakraborty et al., 2020; Mücke et al., 2023; Otgonbaatar & Datcu, 2021), decision tree splitting (Yawata et al., 2022), Restricted Boltzmann Machines (Xu & Oates, 2021), and Deep Belief Networks (Date et al., 2021), can be converted into a QUBO model. The process of transforming optimization problems into QUBO format generally involves: reformulating inequality constraints as equalities using slack variables; converting equality constraints into quadratic penalty terms; representing numerical variables (discrete and continuous) with finite sets of permissible values; and finally, encoding these values using binary schemes to finalize the QUBO formulation (Au-Yeung et al., 2023). A mathematical demonstration of converting SVM to QUBO is available in Appendix 2 of the original document. AQML methodologies have found use in numerous domains, such as materials development candidate selection (Guan et al., 2021; Hatakeyama-Sato et al., 2022; von Lilienfeld, 2018), traffic scheduling optimization (Daugherty et al., 2019), remote sensing data classification (Cavallaro et al., 2020; Delilbasic et al., 2021), anomaly detection (Liu & Rebentrost, 2018), sensor data processing and quantum walks in robotics (Petschnigg et al., 2019), and improving predictions in renewable energy development (Ajagekar & You, 2022).

AQML also finds applications in business areas like financial fraud detection (Grossi et al., 2022; H. Wang et al., 2022), workforce assignment in production planning (Wang et al., 2020), the traveling salesman problem in distribution (Helsgaun, 2007; Ma et al., 2016), and parallel machine scheduling, (Wang and Alidaee, 2019). Computational constraints in implementing AQML models primarily stem from their intensive training requirements. To overcome this fundamental challenge, researchers have deployed approximation techniques like Simulated Annealing (SA) (D-Wave Inc, 2021a) and Multiple Start Tabu Search (MSTS) (D-Wave Inc, 2021b)—both quantum annealing heuristics—which significantly accelerate the QUBO-based AQML training pipeline through more efficient optimization approaches.

However, existing quantum annealing solvers face scalability challenges with large QUBO problems, partly due to the Python wrapper implementations. For such large instances, these solvers often need to break down the QUBO data into smaller sub-problems, solve them individually, and then reassemble the solutions, which can be computationally inefficient due to this "divide and conquer" approach. These scalability issues motivate the creation of faster local search algorithms to enhance the solution quality of QUBO problems. Before introducing the hybrid heuristic, the paper provides theoretical proofs, along with a sufficient and necessary requirement that, once a 1-flip search achieves local optimality, it enables a significant reduction in the number of candidate r-flip moves. The QUBO problem can be mathematically expressed as:

$$Max\ f(x) = \sum_{i=1}^{n} q_i x_i + \frac{1}{2}\sum_{i=1}^{n}\sum_{j\neq i}^{n} q_{i,j} x_i x_j, \ s.t.\ x_i \in \{0,1\}, i = 1,\cdots,n \tag{1}$$

## 2. Previous Works

Developing closed-form expressions for r-flip moves has advantages in creating algorithms to solve very large-scale problems by reducing the computation time of an algorithm. A number of theorems that provide r-flip formulas in closed form for generic pseudo-boolean optimization are presented by Alidaee et al. (2010). Specifically, Theorem 6 in their work pertains to the $f(x) = x^T Q x$ problem. In order to understand the closed-form r-flip formula in this context, some definitions are first introduced.

With respect to a solution vector *x*, the derivative of *f(x)* is given by Equation 3:

$$E(x_i) = q_i + \sum_{j<i} q_{j,i} x_j + \sum_{j>i} q_{i,j} x_j, \ i = 1,\cdots,n \tag{3}$$

**Fact 1.** With respect to $x = (x_1, \cdots, x_n)$, and a new $x' = (x_1, \cdots, 1 - x_i, \cdots, x_n)$ derived by flipping the $i^{th}$ factor of the solution vector, the objective function change is:

$$\Delta f = f(x') - f(x) = (x_i' - x_i)E(x_i) \qquad (4)$$

It is widely recognized that locally optimal solutions for a QUBO problem under a 1-flip must meet the following condition:

$$Either\ ((x_i = 0)\ if\ E(x_i) \leq 0)\ or\ (x_i = 1\ if\ E(x_i) \geq 0),\ for\ i = 1, \cdots, n \qquad (5)$$

Moreover, when $x$ is changed to $x'$, the update for $E(x_j)$, $j=1,...,n$, is determined in (6):

$$\begin{aligned}
&\forall j < i, E(x_j) \leftarrow E(x_j) + q_{j,i}(x_i' - x_i) \\
&\forall j > i, E(x_j) \leftarrow E(x_j) + q_{i,j}(x_i' - x_i) \\
&j = i,\ E(x_j) \leftarrow E(x_j)\ \text{(This implies the contribution of E(x}_i\text{) to } \Delta f \text{ flips sign if E(x}_i\text{) was}
\end{aligned} \qquad (6)$$

used, or that $E(x_i)$ needs re-evaluation for the new $x_i$ state, or simply that this term is regarding the old $E(x_i)$ for delta calculation)

The term $x_i' - x_i$ can also be expressed as $1 - 2x_i$, which simplifies implementation. Algorithm 1 outlines a basic 1-flip search. The strategy of processing elements in a sequence, as indicated in line 3 of Algorithm 1, has proven effective (Alidaee & Wang, 2017; Alidaee et al., 2017; Haibo Wang et al., 2020).

Place Algorithm 1 here

The r-flip search has been included to the premise of Fact 1 (Alidaee et al. (2010)). Given that $x$ is a QUBO solution and that an r-flip move is used to extract x' from x (for $S \subseteq N$, $|S|=r$). The objective function change is:

$$\Delta f = f(x') - f(x) = \sum_{i \in S}(x'_i - x_i).E(x_i) + \sum_{i,j \in S, i<j}(x'_i - x_i).(x'_j - x_j)q_{i,j} \quad (7)$$

Changing $x$ to $x'$, Equation 8 can be used to calculate the update for $E(x_j), j=1,...,n$.

$$\forall j \in N \setminus S, E(x_j) \leftarrow E(x_j) + \sum_{i \in S}(x'_i - x_i)q_{i,j}$$
$$\forall j \in S, E(x_j) \leftarrow E(x_j) + \sum_{i \in S \setminus \{j\}}(x'_i - x_i)q_{i,j} \quad (8)$$

As Alidaee et al. (2010) explained, evaluating the change (Eq. 7) can be accomplished in O($r^2$) time. Updating all $n$ variables' derivatives using Eq. 8 necessitates $r$ calculations for each $j$ of N\S and $r-1$ calculations for each j of S, resulting in an overall O(nr) update time for any factors $i$, $j=1,...,n$, and $i<j$, following is defined:

$$E'(x_i) = E(x_i) - q_i - q_{i,j}x_j$$
$$E'(x_j) = E(x_j) - q_j - q_{i,j}x_i \quad (9)$$

Utilizing Eq. (9), Eq. (7) can be expressed as Eq. (10).

$$\Delta f = \sum_{i \in S}\left[(1 - 2x_i)E'(x_i) + \sum_{j \in S, j \leq i}(1 - x_i - x_j)q_{i,j}\right] \quad (10)$$

(Note: The condition j≤i in the second sum of Equation 10 is unusual if $q_{i,j}$ is symmetric and i,j are distinct. If i=j, $q_{i,i}$ is usually part of the linear term. This might be a specific formulation trick from the cited Alidaee et al. 2010 or a typo if E' already accounts for linear terms).

Algorithm 2 provides a straightforward exhaustive r-flip search. The problem's complexity means that using larger r values in r-flip local search can result in a more time-consuming implementation, though larger $r$ might allow searching a more diverse solution space and finding better solutions. More complicated algorithms like VNS, F&F, and MENS frequently use $r=1$ (and $r=2$) as their fundamental building blocks. Finding a local optimal solution regarding a 1-flip search can greatly minimize an $r$-flip search implementation, as demonstrated by **Theorem 1** and **Proposition 1**. Additionally, the findings are provided to facilitate the effective use of an r-flip search in an algorithm.

Place Algorithm 2 here

## 3. Results for Closed-form Formulas

The number of combinations of *m* items from *n* (for m < n,) is shown by (m, n). Let $\varphi = \text{Max}_{i,j \in N}\{|q_{ij}|\}$ and $M = \varphi * (r, 2)$ (i.e., $\varphi * C(r,2)$). Lemma 1 and Lemma 2, presented next, aid in proving the main results. Lemma 1 is a direct consequence of prior findings(Alidaee et al., 2010; Boros et al., 1999).

**Lemma 1.** For a solution $x = (x_1, ..., x_n)$ that is locally optimal in a 1-flip search:

$$(x'_i - x_i)E(x_i) \leq 0, for\ i = 1,\ldots,n.$$

(11)

**Proof.** Local optimality condition (5) implies that ($E(x_i) \leq 0$ if $x_i = 0$) and ($E(x_i) \geq 0$ if $x_i = 1$), which directly leads to Inequality (11).

**Lemma 2.** Let $x = (x_1, ..., x_n)$ represent a solution. Then, for a set S with |S|=r:

$$\sum_{i,j \in S}(x'_i - x_i)(x'_j - x_j)\, q_{i,j} \leq M \qquad (12)$$

where $(x'_i - x_i)(x'_j - x_j)$ is $+1$ or $-1$, and the sum is over distinct pairs

**Proof.** The term (x'$_i$ - x$_i$)(x'$_j$ - x$_j$)q$_{ij}$ can be q$_{ij}$ or -q$_{ij}$ (assuming distinct i, j and symmetry of $q_{i,j}$ leads to C(r,2) terms) for each i, j $\in$ S pair. Since |S|=r, the summation over C(r,2) pairs will be at most M. The proof implies this, as each term is $q_{i,j}$ or $-q_{i,j}$. Summing C(r,2) such terms, the max is C(r,2)*phi = M).

**Theorem 1:** Given a 1-flip search where $x$ is a local optimum solution for $x^T Q x$, an improved r-flip move is a subset $S \subseteq N$ with $|S| = r$ if:

$$\sum_{i \in S} |E(x_i)| \leq \sum_{i,j \in S} (x'_i - x_i)(x'_j - x_j) q_{i,j} \qquad (13)$$

**Proof:** Utilizing Eq. (7), S of r factors is an r-flip that improves when $\Delta f > 0$:

$$\Delta f = f(x') - f(x) = \sum_{i \in S} (x'_i - x_i) E(x_i) + \sum_{i,j \in S} (x'_i - x_i)(x'_j - x_j) q_{i,j} > 0 \qquad (14)$$

Since x is 1-flip locally optimal, Lemma 1 states $(x'_i - x_i) E(x_i) \leq 0$, so $-(x'_i - x_i) E(x_i) = |E(x_i)|$ (as $(x'_i - x_i)$ is +1 if $x_i = 0$ and -1 if $x_i = 1$, and $E(x_i)$ has the 'opposite' sign for optimality). Thus, inequality (14) is equivalent to:

$$\sum_{i,j \in S} (x'_i - x_i)(x'_j - x_j) q_{i,j} > -\sum_{i \in S} (x'_i - x_i) E(x_i) = \sum_{i \in S} |E(x_i)| \qquad (15)$$

This completes the proof.

**Proposition 1:** If $S \subseteq N$ and $|S| = r$ is an r-flip that improves, then $\sum_{i \in S} |E(x_i)| < M$ where $x$ is a 1-flip locally optimal solution.

**Proof:** From Theorem 1 (Inequality 13, effectively 15), for an improving r-flip:

$$\sum_{i \in S} |E(x_i)| < \sum_{i,j \in S} (x'_i - x_i)(x'_j - x_j) q_{i,j} \qquad (16)$$

From Lemma 2, the right-hand side is $\leq M$. Therefore.

$$\sum_{i \in S} |E(x_i)| < \sum_{i,j \in S} (x'_i - x_i)(x'_j - x_j) q_{i,j} \leq M \qquad (17)$$

If a solution is 1-flip locally optimal and no subset $S$ with $|S| = r$ fulfills Inequality (13) condition, then it is r-flip local optimal as well. This is a corollary of Theorem 1. If no subset $S$ of any size fulfills this condition, then the solution is a local optimum for r-flip searches for all r ≤ n. The same findings apply to Proposition 1.

The significance of Proposition 1 lies in its application to implementing an r-flip search. This shows that if an r-flip search is performed following a 1-flip search, only elements whose sum of absolute derivative values is less than M need consideration. This allows for quick checks to determine if an element $x_i$, alone or with others, is eligible for an improving r-flip move $S$ as illustrated in the following example.

**Example 1**. Consider $x$ to be a 1-flip local optimal solution for an $x^T Q x$ problem with $n$ variables. Let $S = \{i, j, k, l\}$ be a potential 4-flip. For S to be an improving move, all 15 inequalities (derived from Proposition 1 for all subsets of S of size 1, 2, 3, and 4) must be satisfied. If the inequality for the full 4-element set holds (sum of their $|E(x\_a)| < M$ for r=4), then it holds for all its subsets too (using appropriately sized M for each subset). This is key for dynamic neighborhood searches considering k-flips up to $r$. The following strategies are two of many strategies available to look for an improving $S$ subset.

*A. Strategy 1*

A set D(n) is defined, containing subsets that can be improved. Considering a 1-flip local optimal solution vector, its elements are ordered by increasing absolute derivative values:

$$|E(x_i)| \leq \ldots \leq |E(x_{\pi(n)})| \tag{18}$$

where π(i) is the i<sup>th</sup> element in this sorted order. Then, for k=1, 2, ..., n, the summation in (19) is checked. Consider K to be the largest $k$ satisfying this inequality. The set D(n) is then $\{x_{\pi(1)}, \ldots, x_{\pi(K)}\}$

$$\sum_{i=1}^{k}|E(x_{\pi(i)})| < M, for\ k = 1,2,\ldots,n \tag{19}$$

$$D(n) = \{x_{\pi(1)}, \ldots, x_{\pi(K)}\} \tag{20}$$

**Lemma 3.** Any subset $S \subseteq D(n)$ meets the requirements for an improved move (Proposition 1).

**Proof.** This is a direct result of Proposition 1.

a) Ordering elements of x as in (18) offers advantages:

Smaller $|E(x_i)|$ values make $x_i$ more likely to be part of an improving k-flip ($k \leq r$), as it helps satisfy Inequality (19) given a constant M.

b) This ordering allows for a straightforward series of candidate subsets S to be considered, such as those in (21):

$$S \in \{\{\pi(1), \pi(2)\}, \{\pi(1), \pi(2), \pi(3)\}, ., \{\pi(1), ., \pi(K)\}\} \tag{21}$$

Many other subsets of D(n) could also be candidates. To help solve the aforementioned issue up to a certain extent, a weaker version of Proposition 1 is introduced in the Appendix as Proposition 2.

*B. Strategy 2*

An alternative efficient strategy involves using individual elements to form a candidate set D(1) for an r-flip search, based on Corollary 1:

$$D(1) = \{x_i : |E(x_i)| < M\} \tag{22}$$

**Corollary 1:** If x is 1-flip locally optimal, and an r-flip move on elements of S ($i \in S \subseteq N$) improving f(x), then for each $i \in S$, we must have $|E(x_i)| < M$

(Note: The original Corollary statement has "i ∈ S with S ⊆ N". The text seems to imply each element *individually* satisfies this if it's part of an improving r-set, not just the sum. This is a stronger condition than Proposition 1 for individual elements if M is the global M for r-flips).

Place Table 1 Here

Experiments are conducted to determine the size of D (1) for various instances to gain an understanding of how to apply Corollary 1. A procedure is provided in the supplemental materials showing the stages that are part of the experiment. Many researchers use problems that are obtained from the literature (Palubeckis, 2004). Only the larger-scale problems (up to 38 instances) with 2500 to 6000 variables are used.

Results (from the description of Table 1) indicate that as Q density increases, |D (1)| decreases. As problem size increases, |D(1)| also increases. Interestingly, D(1) was smaller for better local optima, suggesting r-flip search time decreases as the global optimum is approached using Corollary 1. Two strategies are implemented as Algorithms 3 and 4, respectively. Algorithm 5 then embeds Strategy 2 within a simple tabu search for improvement.

Place Algorithm 3 here
Place Algorithm 4 here
Place Algorithm 5 here

The procedures Destruction(), Construction(), and randChange() in Algorithm 5 are common in tabu search (Glover et al., 1998; Alidaee and Wang, 2017).

Any local search (Algorithm 3 or 4) can be employed at the initial stage of Algorithm 5. Preliminary tests suggested that Algorithm 4, with slight modification (if 1-flip result is inferior to the current best, stop local search and go to Step 2 of Alg. 5), was effective due to simplicity and low computation time. To compare, Algorithms 3 and 4 (hybrid r-flip/1-flip) are evaluated against Algorithm 5 (hybrid 1-flip/r-flip with tabu search). The first strategy (Algorithm 3) starts with r=1 and dynamically increases r based on Inequality (19). Strategy 2 (Algorithm 4) also uses dynamic r-flips based on Equation (7), increasing r one at a time. The goal is to achieve local optimality on large instances faster. The paper compares these three for 2-flips on very-large-scale QUBOs.

## 4. Computational Results

Extensive computational experiments were conducted to assess the proposed strategies based on the size of the problem, $r$ value, and density. Algorithms 3, 4, and 5 (with $r = 2$) were compared using very large-scale QUBO instances. Consequently, the best of these was then compared to a leading algorithm for $x^T Q x$, the D-Wave tabu solver, which is based on Palubeckis's (2004) Multiple Start Tabu Search (MST2) algorithm. Since the D-Wave solver uses a Python wrapper with size limitations, this study used the original MST2 C++ code (Palubeckis, 2004) for a direct comparison without wrapper-induced limitations.

The hybrid heuristics were coded in C++. Palubeckis's MST algorithms and instance generator (C++) were downloaded. Among Palubeckis's five MST algorithms, MST2 reported the best results and was chosen with default parameters. In MST2, the initial tabu search subroutine uses 25,000 * size_of_problem iterations, then reduces to 10,000 * size_of_problem for subsequent starts. If an improved solution is found within a tabu subroutine, MST2 immediately calls for a local search. Each tabu start subroutine in MST2 ends with checking CPU limits, potentially exceeding short time limits for large instances.

All algorithms ran on a single core of an Intel Xeon Quad-core E5420 (8 GB RAM, 2.5 GHz) using GNU C++ compiler v4.8.5, submitted via Open PBS to ensure consistent CPU and memory usage. Preliminary findings showed Algorithms 3, 4, and 5 performed well on instances < 3,000 variables and low density, finding best-known solutions within 10 seconds. Thus, comparisons focused on large, high-density instances (3,000 - 8,000 variables) against MST2, and the best of Algorithms 3, 4, 5. Several researchers have reported these benchmarks (Glover et al., 2010; Rosenberg et al., 2016). Additionally, very-large-scale QUBOs (30,000 variables, high density) were generated. A 600-second limit for CPU time and r = 2 was used for these three algorithms on these large instances (Table 2).

The results of Algorithms 3, 4, 5 for 30,000-variable instances (10 runs) are provided in Table 2. Algorithm 5 (with r=2) was better than Algorithms 3 and 4. Thus, Algorithm 5 (r = 1 and 2) was compared to MST2, with 60- and 600-second CPU limits per run (10 runs/instance) (Original Paper, p. 21). A tabu tenure (100) was used for 1- and 2-flip in Algorithm 5 and MST2. Instance data is available online at: https://doi.org/10.18738/T8/WDFBR5.

Place Table 2 Here

The stopping criterion for Algorithm 5 (CPU time limit) was checked before starting tabu search, considered fair as neither MST2 nor Algorithm 5 is a single-point based method.

Table 3 details instance size/density, number of times OFV reached, and solution deviation for MST2 and Algorithm 5 (r=1, r=2). MST2 showed consistent performance, frequently reaching the same objective function value. Algorithm 5, starting randomly, explores more diversely in short times. At 600 seconds, both MST2 and Algorithm 5 yielded better solution quality (relative standard deviation, RSD). RSD is calculated as ($100\sigma/\mu$), where $\sigma$ is std. dev. of OFV over 10 runs, and $\mu$ is the mean OFV. For some instances, RSD < 5.0E-4 even if not all runs found same value; 0.000 is used if RSD rounds to this.

Place Table 3 Here

Table 4 presents results for a 60-second CPU limit, Table 5 for 600 seconds. In Table 4 (60s): MST2 matched 5/27 best solutions. Algorithm 5 (1-flip) matched 26/27; Algorithm 5 (2-flip) matched 18/27. MST2 exceeded the 60s limit for initial solution finding in two large instances. In Table 5 (600s), MST2 matched 10/27 best solutions. Algorithm 5 (1-flip) matched 25/27; Algorithm 5 (2-flip) matched 23/27. On high-density large instances, Algorithm 5's 1-flip and 2-flip strategies both demonstrated strong performance. No clear pattern shows 2-flip taking more time than 1-flip for the same objective function value, as initial solutions are chosen independently and randomly. The 1-flip strategy showed better overall performance at both 60 and 600-second limits.

Place Table 4 Here
Place Table 5 Here

Tables 6 and 7 provide the time deviation to reach OFV. MST2 has less variation when it reaches the same objective function value. Algorithm 5 (r-flip) has a wider range of computing times. If OFV is found only once in 10 runs, with zero time deviation.

Place Table 6 Here
Place Table 7 Here

Other local search strategies can incorporate the r-flip technique. Its clever implementation reduces computation time and improves solution quality. Time and solution deviations over 10 runs (short CPU limits) were computed due to resource availability. Compared to MST2 (a key quantum annealing solver component), the r-flip strategy produces high-quality solutions faster and is easily implemented as a "warm start" local search for many quantum annealing solvers.

To compare the performance of three algorithms (MST2, 1-flip, and 2-flip), the initial Friedman's test was conducted (Friedman, 1937). This comparison was based on 27 complete experimental instances, where each instance provided a performance measure (presumably DT) for all three algorithms. The Friedman chi-squared statistic calculated was 19.9808. This statistic reflects the extent to which the rankings of the three algorithms differed consistently across the 27 experiments. A larger statistic suggests more substantial and consistent differences in these rankings. The associated p-value for this test was extremely small, 4.5839e-05 (which is 0.000045839). This p-value represents the probability of observing such a large chi-squared statistic (or larger) purely by chance if, in reality, there were no actual differences in the median performance of the three algorithms (this is the null hypothesis). Since this p-value (4.5839e-05 ) is considerably less than the chosen significance level (alpha) of 0.05, the null hypothesis is rejected. This leads to the conclusion that there is a statistically significant difference in performance among at least two of the algorithms MST2, 1-flip, and 2-flip. However, Friedman's test itself is an omnibus test, meaning it only tells us that a difference exists somewhere among the groups, not specifically which algorithms differ from each other. While Friedman's test itself doesn't pinpoint '2-flip' as statistically superior to both others in all pairwise comparisons, it confirms that the overall pattern of performance, likely influenced by '2-flip's' tendency towards more favorable (lower) DT scores, varies significantly across the three algorithms.

Following the significant result from the Friedman test, Nemenyi's post-hoc test was performed to identify which specific pairs of algorithms exhibited statistically significant performance differences.

Table 8. Nemenyi's Post-Hoc Test Results (pairwise p-values)

|        | MST2     | 1-flip   | 2-flip   |
|--------|----------|----------|----------|
| MST2   |          | 0.000176 | 0.001172 |
| 1-flip | 0.000176 |          | 0.882488 |
| 2-flip | 0.001172 | 0.882488 |          |

Note: A p-value < 0.05 indicates a significant difference between that pair of algorithms. Lower p-values indicate stronger evidence of a difference.

Nemenyi's test provides p-values for each pairwise comparison (Nemenyi, 1963). For the pair MST2 versus 1-flip, the p-value was 0.000176. Since this is less than the alpha of 0.05, we conclude there is a statistically significant difference in performance between MST2 and 1-flip. Similarly, for MST2 versus 2-flip, the p-value was 0.001172, which is also less than 0.05, indicating a statistically significant difference between MST2 and 2-flip. However, when comparing 1-flip versus 2-flip, Nemenyi's test yielded a p-value of 0.882488. This p-value is much greater than 0.05, leading to the conclusion that, according to Nemenyi's test, there is no statistically significant difference in performance between the 1-flip and 2-flip algorithms. Thus, Nemenyi's test suggests that MST2's performance is distinct from both 1-flip and 2-flip, while 1-flip and 2-flip are not demonstrably different from each other in terms of statistical significance. Given that lower DT is better, and assuming MST2 had higher DT values, this would imply MST2 performed significantly worse than both 1-flip and 2-flip.

The Pairwise Wilcoxon Test with Bonferroni Correction was also conducted as another post-hoc analysis to further investigate the pairwise differences, with the Bonferroni correction being a method to control the family-wise error rate when making multiple comparisons (Woolson, 2005).

Table 9. Pairwise Wilcoxon Test with Bonferroni Correction (p-values)

|        | MST2     | 1-flip   | 2-flip    |
|--------|----------|----------|-----------|
| MST2   |          | 0.00005  | 0.000011  |
| 1-flip | 0.000050 |          | 1.000000  |
| 2-flip | 0.000011 | 1.00000  |           |

Note: A p-value < 0.05 indicates a significant difference between that pair of algorithms after Bonferroni correction. Lower p-values indicate stronger evidence of a difference.

The results from this test were highly consistent with Nemenyi's test, though generally producing even smaller p-values for the significant differences, which can happen if the underlying assumptions of Wilcoxon are well met or if it's inherently more powerful for this specific dataset structure. The comparison between MST2 and 1-flip yielded a p-value of 0.000050 (or 5.0e-05), which is well below the 0.05 alpha, confirming a statistically significant difference. The comparison between MST2 and 2-flip produced an even smaller p-value of 0.000011 (or 1.1e-05), also strongly indicating a significant difference. For the crucial comparison between 1-flip and 2-flip, the p-value after Bonferroni correction was 1.000000. This p-value, being far greater than 0.05, reinforces the finding from the Nemenyi test: there is no statistically significant difference in performance between the 1-flip and 2-flip algorithms based on this more conservative post-hoc analysis. Therefore, both post-hoc tests concur that MST2's performance is significantly different from both 1-flip and 2-flip. Since "2-flip (DT) is the lowest, which is the best," and its performance is not statistically distinguishable from 1-flip (which is also significantly different from MST2), this implies that both 1-flip and 2-flip significantly outperform MST2. While 2-flip might have numerically lower average DT values than 1-flip, this difference was not statistically significant enough to be detected by these tests with the current data and significance level.

The proposed algorithms, while demonstrating significant improvements, inherently face several limitations. As heuristic methods, they do not guarantee finding the global optimum, with Algorithms 1, 3, and 4 being susceptible to local optima, a challenge Algorithm 5 (Tabu Search) mitigates but doesn't eliminate. The computational cost of r-flips, though reduced by theoretical pruning (Theorem 1, Proposition 1), still grows with r, making exhaustive checks (Algorithm 2) infeasible for larger problems and rendering even strategic r-flips potentially costly if r is large or derivative updates ($O(nr)$) become bottlenecks. Performance is also sensitive to parameter tuning, including the r-value, tabu tenure, the M-value threshold (dependent on problem coefficients), various probabilistic choices, and iteration counts, all of which require careful setting. The complexity and precise implementation of derivative update rules are crucial; errors can lead to suboptimal search. Furthermore, algorithm efficacy can vary with QUBO instance characteristics (size, density, coefficient distribution), and reliance on CPU time or lack of improvement for stopping might be premature. While these heuristics can serve as effective "warm starts" or standalone methods, their success depends on effectively navigating these limitations and the underlying assumptions, such as achieving a good 1-flip optimum before applying r-flip pruning strategies.

## 5. Conclusion

This paper addressed the QUBO problem, presenting a number of findings such as a condition for r-flip local optimality when a 1-flip search has already converged. This can enhance quantum machine learning models based on QUBO by improving speed and solution quality. Computational results demonstrate a significantly smaller number of candidates for r-flip implementations. Furthermore, the novel r-flip strategy is capable of solving very-large-scale QUBO instances in 600 seconds, using significantly less time to find the best-known solutions for the benchmark problems compared to MST2 (a core algorithm for quantum annealing solvers). These results are promising for very-large-scale problems or sparse matrices in QUBO formulations of quantum machine learning models like quantum SVM, quantum k-means, and quantum feature selection, especially where variable neighborhood strategies are employed.

D-Wave Inc. (2021a). *dwave-SA*. Retrieved Nov 10 from https://docs.ocean.dwavesys.com/projects/neal/en/latest/

D-Wave Inc. (2021b). *dwave-tabu*. Retrieved Nov 10 from https://docs.ocean.dwavesys.com/projects/tabu/en/latest/index.html

Date, P., Arthur, D., & Pusey-Nazzaro, L. (2021). QUBO formulations for training machine learning models. *Scientific Reports*, *11*(1), 10029. https://doi.org/10.1038/s41598-021-89461-4

Date, P., & Potok, T. (2021). Adiabatic quantum linear regression. *Scientific Reports*, *11*(1), 21905. https://doi.org/10.1038/s41598-021-01445-6

Daugherty, G., Reveliotis, S., & Mohler, G. (2019). Optimized Multiagent Routing for a Class of Guidepath-Based Transport Systems. *IEEE Transactions on Automation Science and Engineering*, *16*(1), 363-381. https://doi.org/10.1109/TASE.2018.2798630

Delilbasic, A., Cavallaro, G., Willsch, M., Melgani, F., Riedel, M., & Michielsen, K. (2021, 11-16 July 2021). Quantum Support Vector Machine Algorithms for Remote Sensing Data Classification. (Ed.),^(Eds.). 2021 IEEE International Geoscience and Remote Sensing Symposium IGARSS.

Friedman, M. (1937). The Use of Ranks to Avoid the Assumption of Normality Implicit in the Analysis of Variance. *Journal of the American Statistical Association*, 32(200), 675-701. doi:10.2307/2279372

Glen, S. (2014). *Relative Standard Deviation: Definition & Formula* https://www.statisticshowto.com/relative-standard-deviation/

Glover, F. (1998). A template for scatter search and path relinking. In J.-K. Hao, E. Lutton, E. Ronald, M. Schoenauer, & D. Snyers (Ed.),^(Eds.), *Artificial Evolution*. Berlin, Heidelberg.

Glover, F., Kochenberger, G., & Du, Y. (2019). Quantum Bridge Analytics I: a tutorial on formulating and using QUBO models. *4OR*, *17*(4), 335-371. https://doi.org/10.1007/s10288-019-00424-y

Glover, F., Kochenberger, G. A., & Alidaee, B. (1998). Adaptive Memory Tabu Search for Binary Quadratic Programs. *Management Science*, *44*(3), 336-345. https://doi.org/10.1287/mnsc.44.3.336

Glover, F., Lewis, M., & Kochenberger, G. (2018). Logical and inequality implications for reducing the size and difficulty of quadratic unconstrained binary optimization problems. *European Journal of Operational Research*, *265*(3), 829-842. https://doi.org/https://doi.org/10.1016/j.ejor.2017.08.025

Glover, F., Lü, Z., & Hao, J.-K. (2010). Diversification-driven tabu search for unconstrained binary quadratic problems. *4OR*, *8*(3), 239-253. https://doi.org/10.1007/s10288-009-0115-y20